\documentclass[aps,epsfig,graphics,showpacs,floatfix,twocolumn,mathbbm,nopacs]{revtex4}

\usepackage{amsmath,amsfonts,amssymb,amsthm,stmaryrd,graphics,graphicx,epsfig,color,times,bbm,psfrag}

\usepackage[all]{xy}

\newcommand{\xnode}[1]{*+[F]{#1}}
\newcommand{\xmnode}[1]{*+[F]{A[#1]}}
\newcommand{\xminid}[1]{
	\xymatrix@C=4mm{#1}
}

\begin{document}

\bibliographystyle{apsrev}

\newcommand{\tr}{\operatorname{tr}}
\newcommand{\uinvnorm}{|\kern-2pt|\kern-2pt|}
\newcommand{\wt}{\operatorname{wt}}
\newcommand{\spectrum}{\operatorname{sp}}
\newcommand{\erf}{\operatorname{erf}}
\newcommand{\erfc}{\operatorname{erfc}}
\newcommand{\supp}{\operatorname{supp}}
\newcommand{\diam}{\operatorname{diam}}

\bibliographystyle{apsrev}

\newcommand{\me}{\mathrm{e}}
\newcommand{\mi}{\mathrm{i}}
\newcommand{\md}{\mathrm{d}}

\newcommand{\cc}{\mathbb{C}}
\newcommand{\nn}{\mathbb{N}}
\newcommand{\rr}{\mathbb{R}}
\newcommand{\zz}{\mathbb{Z}}
\newcommand{\id}{\mathbbm{1}}

\newtheorem{definition}{Definition}
\newtheorem{theorem}{Theorem}
\newtheorem{lemma}{Lemma}
\newtheorem{corollary}{Corollary}
\newtheorem{property}{Property}
\newtheorem{proposition}{Proposition}
\newtheorem{remark}{Remark}
\newtheorem{example}{Example}
\newtheorem{assumption}{Assumption}

\setlength{\parskip}{2pt}

\newcommand{\identity}{\openone}
\newcommand{\be}{\begin{equation*}}
\newcommand{\bea}{\begin{eqnarray}}
\newcommand{\eea}{\end{eqnarray}}
\newcommand{\ee}{\end{equation*}}
\newcommand{\bra}[1]{\mbox{$\langle #1 |$}}
\newcommand{\ket}[1]{\mbox{$| #1 \rangle$}}
\newcommand{\braket}[2]{\mbox{$\langle #1  | #2 \rangle$}}
\newcommand{\proj}[1]{\mbox{$|#1\rangle \!\langle #1 |$}}
\newcommand{\ev}[1]{\mbox{$\langle #1 \rangle$}}
\def\sign{\mbox{sgn}}
\def\H{{\cal H}}
\def\C{{\cal C}}
\def\E{{\cal E}}
\def\O{{\cal O}}
\def\B{{\cal B}}
\def\one{\ensuremath{\hbox{$\mathrm I$\kern-.6em$\mathrm 1$}}}
%{\ensuremath{\hbox{$\mathrm I$\kern-.6em$\mit 1$}}}
\def\tr{ \mbox{tr}}

\newcommand{\note}[1]{{\sc #1}}
\newcommand{\diff}{}

\bibliographystyle{unsrt}

\title{Novel schemes for measurement-based quantum computation}

\author{D.\ Gross and J.\ Eisert}

\affiliation{
1 Blackett Laboratory, 
Imperial College London,
Prince Consort Road, London SW7 2BW, UK\\
2 Institute for Mathematical Sciences, Imperial College London,
Exhibition Rd, London SW7 2BW, UK}

\date\today

\begin{abstract}
We establish a framework which allows one to 
construct novel schemes for measurement-based quantum computation. 
The technique further develops tools from many-body physics -- 
based on finitely correlated or projected entangled pair states -- 
to go beyond the cluster-state based one-way computer.
We identify resource states that are radically
different from the cluster state, in that they exhibit
non-vanishing correlation functions, can partly be prepared
using gates with non-maximal entangling power, or 
have very different local entanglement properties. In the
computational models, the randomness is compensated
in a different manner. It is shown that there exist 
resource states which are locally arbitrarily close to a 
pure state. Finally, we comment
on the possibility of tailoring computational models
to specific physical systems as, e.g.\ cold atoms in optical lattices.
\end{abstract}


\maketitle

No classical method is known which is capable
of efficiently simulating the results of measurements 
on a general many-body quantum system:
the exponentially large state space renders 
this a tremendously difficult 
task. What is a burden to computational physics 
can be made a virtue in quantum information science:
It has been shown
that multi-particle quantum states can form resources for quantum
computing \cite{Computers}.  
Indeed, universal quantum computation is
possible by first preparing a certain multi-partite 
entangled resource  -- 
called a \emph{cluster state} \cite{Cluster}, which does not 
depend on the algorithm to be implemented -- 
followed by local measurements on the 
constituents. This idea of a
measurement-based ``one-way computer'' ($QC_{\cal C}$) \cite{Oneway} has
attracted considerable attention in recent years.
Progress has indeed been made concerning a 
systematic understanding of the computational model 
of the one-way computer as such 
\cite{Survey,CompSurvey,GS,Vedral,vbqi}. 
Quite surprisingly, this contrasts with the lack of 
development of new
computational models or novel resource states
beyond that original framework. To our knowledge -- no single model distinct from the $QC_{\cal C}$ has been
developed based on local measurements on a fixed,
algorithm-independent qubit resource state.  Hence, questions of
salient interest seem to be: {\it Can we systematically find
alternative schemes for measurement-based quantum computation?  What are the properties that distinguish computationally universal resource
states?} 

These questions are clearly central when thinking of tailoring
resource states to specific physical systems, e.g.\ to cold atoms in
optical lattices, purely linear optical systems or condensed-matter
ground states. They are also of key interest when addressing the
question what flexibility one has in the construction of such schemes,
and what properties of may ultimately be relaxed.  The problem is also
relevant to many-body physics, when the question of efficient
classical simulatability \cite{vidal} is addressed: Quantum states may
thought of being ordered according to their computational potency,
universal and efficiently simulatable states forming the respective
extremes.

In this work, we demonstrate how methods 
from many-body physics can be
extended to develop schemes for measurement-based quantum
computation (MBC). Starting from 
the concepts of matrix-product,
projected entangled pair, and 
finitely correlated states \cite{PEPS,FCS}, 
we develop a framework broad enough 
to allow for the construction of novel universal resources
and models. 
The notion of universality in the context of one-way computing 
was recently addressed in Ref.\ \cite{maarten1}.  A \emph{universal
resource} in their sense is a family of states out of which any other
state can be obtained by local measurements on a subset of sites. 
It follows from the definition that many states cannot be universal:
E.g.\ states which are locally non-maximally entangled, have
non-vanishing two point correlation functions $\langle O_i O_{i+r}
\rangle - \langle O_i \rangle\,\langle O_{i+r}\rangle$ or a
non-maximal localizable entanglement \cite{maarten1}. 
Complementary to this approach, we refer to a device as a 
universal quantum computer, if it can efficiently predict the
outcome of any quantum algorithm. A state will hence be called a
universal resource if one can, assisted by the results of local
measurements on the state, efficiently predict the result of any
quantum computation.

\begin{figure}[!h] \label{fig:resources} 
 \begin{center}
      \leavevmode
\resizebox{8.8 cm}{!}{\includegraphics{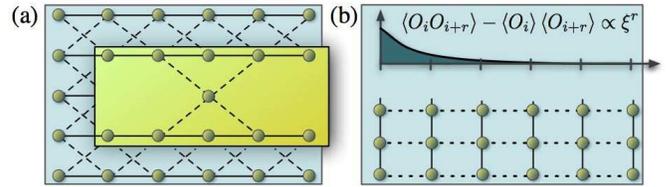}}
\end{center}
\caption{Two resources
for universal measurement-based 
quantum computing. Fig.\ 1a 
depicts a weighted graph state, where 
solid lines correspond to a controlled $\pi$-phase gate, 
dashed lines to $\pi/2$. Fig.\ 1b
represents a scheme deriving from an AKLT-type model.
Dashed lines represent a state with 
non-vanishing correlation functions, solid lines 
correspond to $\pi$-phase gates in the $\{\ket1,\ket2\}$-subspace.}
\end{figure}

To exemplify the power of our framework,
we describe three new models for MBC
in quantum mechanical lattice systems. In
all these models, the randomness is compensated 
in a manner different from the $QC_{\cal C}$. 
They are chosen to highlight that, intriguingly, many properties of the original 
one-way computer may be relaxed, 
while still retaining a 
universal model for quantum computing: (i) We find 
resources exhibiting non-vanishing two-point correlations (which are
typical for natural many-body ground states). 
The original discussion of the $QC_\mathcal{C}$ 
depended on the fact 
that the cluster can be prepared by mutually commuting unitaries
(technically a quantum cellular automaton (QCA) \cite{QCA}).
Commutativity enables one to logically break down a $QC_{\cal
C}$-calculation into small parts corresponding to individual gates;
however, it implies severe restrictions, such as that the correlations
vanish completely outside some neighborhood. 
Hence our framework can prove universality for states 
not amenable to any QCA-based
technique. 
(ii) 
We treat a universal weighted graph state with partly weakly entangled
bonds.
(iii) 
In the final part -- using different techniques -- we 
present a family of states which
are universal, yet are locally arbitrarily close to a pure state.

{\it Matrix product states. -- } Starting point
in the 1-D setting is the familiar notion of a matrix 
product  state (MPS) \cite{FCS}.
%; the difference arising from 
%the treatment of flow of quantum information under measurements.
We will first look at the simple case of a chain of $n$ qubits.
Its state is specified by (i) an 
auxiliary $D$-dimensional Hilbert space, called
\emph{correlation space}, 
(ii) two operators $A[0], A[1]$ on $\cc^D$,
and (iii) two vectors $\ket L, \ket R$ representing boundary
conditions. One has explicitly
\begin{equation}\label{eqn:linearMps}
	 \ket\Psi=\sum_{s_1,\dots, s_n=0,1} \bra R A[s_n] \dots A[s_1]
	 \ket L \,\,\ket{s_1, \dots, s_n}. 
\end{equation}
In order to generalize Eq.\ (\ref{eqn:linearMps}) to 2-D lattices,
we need to cast it into the form of a tensor network. 
Setting 
$
	L_i=\braket iL, 
	A[s]_{i,j}=\bra j A \ket i,
$ 
we arrive at
$
	\langle s_1, \dots , s_n | \Psi \rangle = 
	\sum_{i_0,\dots,i_n}^D
	L_{i_0} 
	A[s_1]_{i_0,i_1} \dots A[s_n]_{i_{n-1}, i_n} 
	{R^\dagger}_{i_n}
$.

{\it Computational tensor networks. --} 
While the 1-D setting is awkward enough, the 2-D equivalent
is completely unintelligible. 
To cure this problem, we introduce a
graphical notation \cite{Birdtrack}
which enables an intuitive 
understanding beyond the 1-D case. 
Tensors will be represented by boxes, indices by
edges:
\begin{equation*}
		L_r=\xymatrix@C=5mm{*+[F]{L}\ar@{-}[r]&},\,
	  A[s]_{l,r}=\xymatrix@C=5mm{\ar@{-}[r]&*+[F]{A[s]}\ar@{-}[r]&},\,
	  {R^\dagger}_l=\xymatrix@C=5mm{\ar@{-}[r]&*+[F]{R^\dagger}}.
\end{equation*}
A single-index tensor can be interpreted as the expansion coefficients
of either a ``ket'' or a ``bra''. Sometimes, we will indicate what
interpretation we have in mind by placing arrows on the edges:
outgoing arrows designating ``kets'', incoming arrows ``bras''.
Connected arrows designate contractions of the respective indices.
%Eq.\ (\ref{eqn:linearMps}) now reads
%\begin{equation*}
%	\braket{s_1\dots s_n}{\Psi}=
%	\xymatrix@C=3mm{
%		\xnode{L}\ar[r]&*+[F]{A[s_1]}\ar[r]&\dots\ar[r]&*+[F]{A[s_n]}
%		\ar[r]&*+[F]{R^\dagger}
%	}.
%\end{equation*}
If $\ket\phi$ is a general state vector in $\cc^2$, we abbreviate
$\braket\phi0\,A[0]+\braket\phi1\,A[1]$ by $A[\phi]$. 
The overlap of $\ket\Psi$ with a local
projection operator  is easily derived:
\begin{equation}\label{eqn:transport}
	\big(\bigotimes_i^n \bra{\phi_i}\big) \ket\Psi 
	= 
	\xymatrix@C=1mm@!{
		*+[F]{L}\ar[r]&*+[F]{A[\phi_1]}\ar[r]&
		\dots\ar[r]&*+[F]{A[\phi_n]}\ar[r]&*+[F]{R^\dagger}
	}.
\end{equation}
Eq.\ (\ref{eqn:transport})
should be read as follows: Initially, the correlation
system is in the state $\ket L$. Subsequent measurements of local
observables with eigenvectors $\ket{\phi_i}$ at the $i$-th site induce
the evolution $A[\phi_i]$, thereby ``processing'' the state in the
correlation space. The probability of a certain sequence of
measurements to occur is given by the overlap of the resulting state
vector with $\ket R$.  An appealing perspective on MBC suggests
itself: Measurement-based computing takes place in
correlation space; the gates acting on the correlation systems are
determined by local measurements.  
The crucial new insight compared to previous treatments of MPS and
PEPS in the context of many-body physics \cite{FCS,PEPS} or MBC
\cite{vbqi} is that 
\emph{the matrices used in the parametrization of an MPS
can be directly understood as quantum gates on a logical space}.  We
will refer to this representation of MBC, 
%going beyond
%projected entangled pair 
%states forming the resources in the treatment of 
%measurements, 
as a \emph{computational tensor network} (CTN).
%Intuitively, ``quantum
%correlations'' are the source of a resource's computational potency.
%The strength of this framework lies in the fact that it assigns a
%concrete mathematical object to these correlations. 
%%It should intuitively be clear how this framework can be 
%%used to trace the flow of information. 

The graphical notation greatly facilitates the passage to 2-D
lattices. Here, the tensors $A[0/1]$ have four indices
$A[s]_{l,d,r,u}$, which will be contracted with the indices of the
left, right, upper and lower neighboring tensors respectively:
\begin{eqnarray} \label{eqn:2d} \braket{s_{1,1}, \dots,
s_{2,2}}{\Psi}= \begin{xy} *!C\xybox{\xymatrix@C=3mm@R=2mm{ &
\xnode{U}        & \xnode{U}     \\
%	 \xnode{L}\ar[r]&\xnode{s_{1,1}}\ar[u]\ar[r]&\xnode{s_{2,1}}\ar[u]\ar[r]&\xnode{R}
%	 \\
%	 \xnode{L}\ar[r]&\xnode{s_{1,2}}\ar[u]\ar[r]&\xnode{s_{2,2}}\ar[u]\ar[r]&\xnode{R}
%	 \\
	 \xnode{L}\ar[r]&\xnode{A[s_{1,1}]}\ar[u]\ar[r]&\xnode{A[s_{2,1}]}\ar[u]\ar[r]&\xnode{R}
	 \\
	 \xnode{L}\ar[r]&\xnode{A[s_{1,2}]}\ar[u]\ar[r]&\xnode{A[s_{2,2}]}\ar[u]\ar[r]&\xnode{R}
	 \\ &\xnode{D}\ar[u]      &\xnode{D}\ar[u]     }} \end{xy}
	 \end{eqnarray} for various boundary conditions $L,D,R,U$. 
Notably, it is known \cite{PEPS} that classical computers cannot efficiently perform
the contraction appearing in Eq.\ (\ref{eqn:2d}). This fact is an
incarnation of the power of quantum computers and no problem to our
approach.
We will now describe several examples, demonstrating the versatility
of our framework and showing how -- surprisingly -- many reasonable
assumptions about universal resources turn out to be unnecessary.  In
what follows, we use the standard notation $X,Y,Z$ for the Pauli
operators, $H$ for the Hadamard gate and $S=\operatorname{diag}(1,i)$
for the $\pi/4$-\emph{gate}. The \emph{controlled $\phi$-phase gate}
is $|0,0\rangle\langle 0,0 | + |0,1\rangle\langle 0,1|+
|1,0\rangle\langle 1,0| + e^{i\phi} |1,1\rangle\langle 1,1|$.  
Lastly, $\ket\pm= 2^{-1/2} (\ket0\pm\ket1)$.

{\it Two-point correlations. --} Here we consider a resource with
exponentially decaying correlation functions, in a way as it
occurs in ground states, but not in states resulting from a QCA. 
To be brief, we first describe a 1-D setting, turning 
to 2-D structures later. Define $G:=\exp(i \pi/k X)$ for some integer
$k>2$.
The relations
\begin{eqnarray*}
	&
	\xymatrix@C=4mm{
		\ar[r]&\xnode{A[s]}\ar[r]&
	}=G \ket s_r \bra s_l,\quad
	\ket L = \ket +,\quad \ket R = G^{-1} \ket +
	&
\end{eqnarray*}
define a state vector $\ket\Psi$ for a chain of qubits. 
The two-point correlations never vanish completely: One finds 
\cite{FCS} that 
\begin{equation*}
\langle Z_i Z_{i+n+1} \rangle-\langle Z_i\rangle\,\langle
	Z_{i+n+1}\rangle \propto \xi^n, 
\end{equation*}	
where
	$\xi=2\sin^2(\pi/k)-1$.
Still, all single-qubit unitaries 
on the correlation system can be realized by local physical
measurements. 
Ignoring global factors (as we will do when possible), one
computes:
%\begin{equation}\label{eqn:esTransport}
%	\xminid{\ar[r]&\xnode{A[+]}\ar[r]&} = G, \quad 
%	\xminid{\ar[r]&\xnode{A[-]}\ar[r]&} = GZ.
%\end{equation}
%Eq.\ (\ref{eqn:esTransport}) will be represented more compactly as 
%\begin{equation}\label{eqn:boxedTransport}
%	\xminid{\ar[r]&\xnode{A[X]}\ar[r]&}=G Z^x,
%\end{equation}
%so an observable as the argument to $A[\,]$ denotes a 
%measurement in
%the corresponding eigenbasis. 
\begin{equation*}\label{eqn:esTransport}
	\hspace{-3mm}
	\xymatrix@C=3mm{\ar[r]&\xnode{A[+]}\ar[r]&}
	\hspace{-3mm}=G,
	\hspace{-2mm}
	\xminid{\ar[r]&\xnode{A[-]}\ar[r]&}
	\hspace{-3mm}=GZ;\>
	\xminid{\ar[r]&\xnode{A[X]}\ar[r]&}\!\!\!\!\!=G Z^x,
\end{equation*}
where the r.h.s. is a compact notation for the two equations on the
left:
An observable as the argument to $A[\,]$ denotes a 
measurement in the corresponding eigenbasis. 
The outcome of the measurement is
assigned to a variable; here $x=0$ in case of the $+1$-eigenvalue and
$x=1$ in case of $-1$. Local
$X$-basis measurements hence 
cause the state of the  correlation system to be
transported from left to right (up to local unitaries).
When measuring several consecutive sites in the $X$-basis, the overall
operator applied to the correlation system
is given by $B := \dots G Z^{x_2} G Z^{x_1}$.
Assuming that we intended to just transport the information
faithfully, we
conceive $B$ as an unwanted \emph{by-product}. To
understand this structure, consider the following 
elementary statement:
	Let $A,B$ be matrices having finite order \cite{Finite}. 
	Every
	element in the group generated by $A,B$ can be written as
	$A B^{k_1} A B^{k_2} A \dots A B^{k_n} $
	for some $n\in \nn$ and $k_i \in \{0,1\}$.
Applied to our situation: 
The by-product operators form a finite group
$\mathcal{B}$ generated by $G,Z$.
The group property gives a possibility to cope
with by-products \cite{Disclaimer}: Assume that at some point the
state vector of the correlation system is given by $B \ket\psi$, for some
unwanted $B\in \mathcal{B}$. Transferring the state along the chain
will introduce any by-product $B'\in\mathcal{B}$ after a finite
expected number of steps. In particular, $B'=B^{-1}$ will occur,
leaving us with $B^{-1} B \ket\psi=\ket\psi$. Note that this technique
is completely general: it can deal with any finite by-product group
(see further examples below). Moving on,
a measurement in the $2^{-1/2}(\ket0
\pm e^{i\phi}\ket1$ basis induces the operator
$S(\phi):=\operatorname{diag}(1,e^{i\phi})$ on the correlation system
(up to by-products). But by the preceding discussion, we can also
implement $GS(\phi)G^{-1}$, which is sufficient to generate any
single-qubit unitary \cite{Computers}.
Lastly, it is easy to see that $Z$-measurements prepare a known state
in the correlation system and conversely can be used to read it out.

{\it AKLT-type states. -- } In this example, we consider
ground states of nearest-neighbor spin-1 
Hamiltonians of the AKLT-type, as they are
well-known in the context of 
condensed-matter physics \cite{FCS}. We 
investigate ground-states induced by
$
	\xminid{\ar[r]&\xnode{A[0]}\ar[r]&} = H$,
	$\xminid{\ar[r]&\xnode{A[1]}\ar[r]&} = 
	(X-iY)2^{-1/2}$,
	$\xminid{\ar[r]&\xnode{A[2]}\ar[r]&} = 
	(X+iY)2^{-1/2}$.
This is the exact unique ground state of a nearest-neighbor 
frustration-free gapped Hamiltonian (in the original AKLT model 
$H$ is replaced by $Z$ in the definition of $A[0]$). Proving 
that any single-qubit unitary can be realized on the correlation 
space commences in a similar way as 
before. Measurements in the $\{\ket0, 2^{-1/2}(\ket1\pm
e^{i\phi}\ket2)\}$-basis gives rise to $H, S(\phi)$ or $Z S(\phi)$,
depending on the measurement outcome. 
The finite by-product group is
in this case generated by $H, Z$. But that is all we need to show, as
gates of the form $S(\phi), HS(\phi)H$ generate all of $SU(2)$.

{\it Weighted graph states. -- } Both previous examples can be embedded
into 2-D lattices, universal for computation (see Fig.\ 1b). A
general technique for coupling 1-D chains to 2-D
universal resources will be discussed by means of a further example: the
\emph{weighted graph state} \cite{Weighted,Survey} shown in Fig.\ 1
(a). In the figure, vertices denote physical systems initially in
$\ket+$, solid edges the application of a controlled $\pi$-phase gate
and dashed edges controlled $\pi/2$-phases, so some of the entangling
gates do {\it not have maximal entangling power}.  The resource's
tensor representation (acting on a $D=2$-dimensional correlation
space) is given by
\begin{equation}
\hspace{-1mm}
\begin{xy}
*!C\xybox{\xymatrix@C=3mm@R=1mm{
								&    & \\
 \ar[r]&\xnode{A[s]}\ar[lu]\ar[r]\ar[ru]& \\
 \ar[ru]&                            &\ar[lu]
}}
\end{xy}
=S^s\ket +_{ru}\,S^s\ket +_{lu}\,Z^s\ket+_r\bra s_{ld}\bra s_{rd}\bra s_l,
\label{eqn:triangleDef}  
\end{equation}
where $s\in\{0,1\}$. Indices
are labeled $ru$ for
``right-up'' to $ld$ for ``left-down''. Boundary conditions are $\ket
0$ for the $ru,lu,r$-directions; $\ket+$ otherwise.
The broad setting for our scheme is the following: the correlation
system of every second horizontal line in the lattice is interpreted
as a logical qubit.  Intermediate lines will either be measured in the
$Z$-eigenbasis -- causing the logical bits to be isolated -- or in the
$Y$-basis -- mediating an interaction between adjacent logical qubits.
%(this is still similar to the $QC_{\cal C}$).

We will first describe how to realize isolated evolutions of logical
qubits. According to Eq.\ (\ref{eqn:triangleDef}) the tensors
$A[0/1]$ factor, allowing us to draw only the arrows corresponding to
the factors of interest; so e.g.\ $\xminid{\xnode{A[s]}\ar[r]&}=Z^s
\ket+_r$.  We find
\begin{equation}\label{eqn:2dto1d}
	\begin{xy}
	*!C\xybox{\xymatrix@C=3mm@R=0mm{
	\xmnode{Z_{i-1,u}}&    &\xmnode{Z_{i+1,u}} \\
					\ar[r]	 &\xmnode{X_i}\ar[lu]\ar[ru]\ar[r]& \\
	\xmnode{Z_{i-1,d}}\ar[ru]&    &\xmnode{Z_{i+1,d}}\ar[lu] \\
	}}
	\end{xy}
	=H S^{2x_i+z_i},
\end{equation}
where $z_i=z_{i-1,u}+z_{i-1,d}+z_{i+1,u}+z_{i+1,d}$. Eq.\
(\ref{eqn:2dto1d}) is of the kind treated before in the case of 1-D
chains. Indeed, using the same techniques, one sees easily that general
local unitaries can be implemented by measurements in the
$2^{-1/2}(\ket0\pm e^{i\phi}\ket1)$ basis. The by-product group
here is given by the full single-qubit Clifford group.
%, generated
%by phase and Hadamard  gates.
%
Turning to two-qubit interactions, consider 
the schematics for a controlled-$Z$ gate
(we suppress adjacent sites measured in the $Z$-basis):
\begin{equation}
	\begin{xy}
	*!C\xybox{\xymatrix@C=5mm@R=1mm{
	\ar[r]&\xmnode{X}\ar[r]& \xmnode{X}\ar[r]& \xmnode{X}\ar[r]& \\
        &          &\xmnode{Y}\ar[lu]\ar[ru]\\
	\ar[r]&\xmnode{X}\ar[r]\ar[ru]&\xmnode{X}\ar[r]&\xmnode{X}\ar[lu]\ar[r]& 
	}}
	\end{xy}.
\end{equation}
In detail: first we perform the $X$-measurements on the sites shown
and the $Z$-measurements on the adjacent ones. If any of these
measurements yields the result ``$1$'', we apply a 
$Z$-measurement to the
central site labeled $Y$ and restart the 
procedure three sites to the
right \cite{Disclaimer}. If all outcomes 
are ``$0$'', a $Y$-measurement is
performed on the central site, obtaining the result $y$.  Let us
analyze the gate step by step. For $c\in\{0,1\}$,
\begin{equation*}
\begin{array}{l}
\begin{xy}
	*!C\xybox{\xymatrix@C=3mm@R=2mm{
	\ar[r]&\xmnode{X}\ar[r]& \xmnode{X}\ar[r]& \xmnode{X}\ar[r]& \\
				& {S^c\ket+}\ar[u] &                 & {S^c\ket+}\ar[u]
	}}
	\end{xy} = H Z^c, \\
\begin{xy}
*!C\xybox{\xymatrix@C=0mm@R=1mm{
								&&&\\
								&\xmnode{Y}\ar[lu]\ar[ru]\\
			 {S^c\ket+}\ar[ru] &                 & {S^c\ket+}\ar[lu]
}}
\end{xy}
\hspace{-5mm}
= (\id + (-1)^{c+y}i S^c\otimes
S^c)\ket+_{lu}\,\ket+_{ru}\\
\begin{xy}
*!C\xybox{\xymatrix@C=3mm@R=2mm{
								&&&\\
	{\ket c}\ar[r]&\xmnode{X}\ar[u]\ar[r]& \xmnode{X}\ar[r]& \xmnode{X}\ar[r]\ar[u]& \\
}}
\end{xy}
\hspace{-2mm}
= S^c \ket+_{lu}S^c \ket +_{ru} H\ket c_r.
\end{array}
\end{equation*}
In summary, the evolution afforded on the upper line is 
$H(\id + (-1)^{c+y} i Z) \propto H S Z^{y+c}$, equivalent to $Z^c$ up to
by-products.
This completes the proof of universality.
For completeness, note that we never need the by-products to vanish
for all logical qubits simultaneously. Hence the expected number of
steps for the realization of one- or two-qubit gates is a constant in
the number of total logical qubits.

%The approach of ``transporting forward until the right outcomes occur'' has
%been chosen for two reasons: firstly, it greatly simplifies the proof.
%Secondly, the large by-product group makes it a necessity: unlike the
%the $QC_{\cal C}$-scheme, not all local by-products are mapped to
%local operators by the logical two-qubit gate. We expect this feature
%to be typical for general MBC methods. Having said that, some elements
%in $\mathcal{B}$ are of course mapped to local operators by the
%controlled-$Z$ operation and this observation (among others) allows one to
%considerably cut down the overhead present in the above argument
%\cite{Prep}. It should be stressed that our emphasize here lies on 
%proofs of principle, as opposed to optimized schemes. 

{\it Entanglement properties of universal resources. --} 
In this section, we further investigate -- using
different methods --  to what extent 
the entanglement properties of the cluster state can be 
relaxed while retaining universality in the above sense.
More specifically, we ask whether one can
find qubit resources that are
(i) universal for $QC_{\cal C}$, 
(ii) translationally invariant, 
(iii) 
%can be prepared from a product state via a QCA, 
which have an arbitrarily small local
entropy and localizable entanglement (LE), and (iv) 
from which not even a Bell pair
can be deterministically distilled?

To show that -- rather surprisingly -- this is indeed the case, we
will encode each logical qubit in blocks of $2k+1$ horizontally
adjacent physical qubits.
Here, $k$ is an arbitrary parameter.  
The first 
$k$ qubits per block
will take the role of ``codewords'', the 
final $k+1$ are ``marker qubits'' used in a construction to make the
resource translationally invariant. 
We start by preparing a regular cluster state in the respective first qubit of
each block.
Then, we encode the states of each of these first qubits 
according to $\ket 0 \mapsto \ket {O_k}:= \ket 0^{\otimes k}$
and
$\ket 1 
\mapsto
\ket {W_k}:=k^{-1/2} (\ket{0,\dots,0,1}+\ket{0,\dots,1,0}+\ket{1,\dots,0,0})$. The rear $k+1$ qubits of each block are 
prepared in $|0,\dots,0,1\rangle$. 
Call the resulting state vector $\ket\phi$.
Finally, the translation invariant resource we will consider
is $|\Psi\rangle = \sum_{t=0}^{2k} {\cal T}^{t} |\phi\rangle$, 
where ${\cal T}$ implements a cyclic translation of the lattice in the
horizontal direction.
To realize universal computing, we 
pick one block and measure each of
its qubits in the  $Z$-basis. 
In this way,  one can deterministically distinguish the 
states ${\cal T}^t
\ket\phi$ corresponding to different values of $t$. 
Indeed, the maker codewords guarantee that a sequence of
$k$ ``0''s followed by a ``1'' appears exactly once in the
result (assuming cyclic boundaries). From the sequence's
position, one easily infers $t$. For definiteness, assume $t=0$.
We then encounter a cluster state
in the encoding $\ket {O_k} $ and $\ket {W_k}$.
The key point to
notice is that, by Ref.\ \cite{Walgate}, any two pure orthogonal
multi-partite states can be deterministically distinguished 
using LOCC. In fact, employing the construction 
of Ref.\ \cite{Walgate}, this can be done by an appropriate 
ordered sequence of adapted projective measurements 
$\pi_1\otimes \dots \otimes \pi_k$ on the sites
of each codeword, the effect corresponding exactly to an
arbitrary given projective dichotomic measurement 
with Kraus operators $|\psi\rangle\langle\psi|$
and $|\psi^\bot \rangle\langle\psi^\bot | = 
\id - |\psi\rangle\langle\psi|$
in the logical space. Hence, 
one can translate any single-site measurement on a cluster state
into an LOCC protocol for the encoded cluster. This shows that
$\ket\Psi$ is universal for deterministic MBC.
At the same time, the von Neumann entropy $S_{vN}$ of any site of the
initial resource is arbitrarily small for sufficiently large $k$: From
the distribution of ``0''s and ``1''s in the codeword, one finds that the
entropy for a measurement in the computational basis reads
$S_Z=H_b(3/(4k+2))$, where $H_b$ is the binary entropy function.
Using the concavity of the entropy function, we have that $LE \leq
S_{vN} \leq S_Z$. It follows that not even a Bell pair can be
deterministically created between any two fixed systems.

{\it Outlook. --} Until now, the only known scheme for MBC was the
$QC_{\cal C}$ and slight variations. Entire classes of states with
physically reasonable properties (e.g.\ non-maximal local entanglement,
long-ranged correlations, weakly entangled bonds)
could not be dealt with. It is our hope that the framework presented
opens up the possibility to adopt the computational model 
to some extent to the specific physical systems at hand and no 
longer vice versa. For example, in linear
optics computing, bonds are the easier to create the lower the
entanglement \cite{ProbaLin}.
Under those circumstances, there may well be a
trade-off between the effort used to prepare a resource and its
efficiency for MBC \cite{ProbaLin}. In turn, for cold atoms in 
optical lattices, exploiting cold collisions \cite{CC}, 
configurations as in  Fig.\ 1 (a) could as feasibly 
be created as the cluster state, making use of a different 
interaction time for diagonal collisions. Other states may well 
be less fragile to finite temperature and decoherence effects. 
The presented tools open up a way for studies of 
quantitatively exploring such trade-offs in preparation. 

%In this work, we have introduced a framework for
%MBC, powerful enough to encompass various different resources and
%computational schemes. Our setting opens up the possibility to think
%about tailor-made resources, suitable for a specific physical
%architecture at hand. For example, in linear optical computing, 
%the success probability of gates is a limiting
%factor.  In this setting, there can be a trade-off
%between the magnitude of the phase of a controlled 
%phase gate and the probability of success \cite{Linear}.  
%Under those circumstances there may well be a
%trade-off between the effort used to prepare a resource and its
%efficiency for MBC \cite{ProbaLin}. In turn, for cold atoms in 
%optical lattices, exploiting cold collisions \cite{CC}, 
%configurations as in  Fig.\ 1 (a) could as feasibly 
%be created as the cluster
%state, making use of a different interaction time for diagonal
%collisions. Other states may well be significantly less
%fragile to finite temperature and decoherence effects. It would
%be interesting to make this trade-off possible by means of 
%this new freedom quantitative in various physical situations.

{\it Acknowledgments. --}  
This work has been supported by the DFG
(SPP 1116), the EU (QAP), 
the EPSRC, the QIP-IRC, 
Microsoft Research, and the EURYI Award Scheme.


\begin{thebibliography}{99}
\bibitem{Birdtrack}
	Similar graphical notations have been used before \cite{graphical}.
\bibitem{Disclaimer}
	This scheme can be vastly
	optimized \cite{Prep}. In the present work, we
	are only interested in proofs of principle.
\end{thebibliography}
\end{document}